\documentclass[fleqn,usenatbib]{mnras}

\usepackage{newtxtext,newtxmath}

\usepackage[T1]{fontenc}

\DeclareRobustCommand{\VAN}[3]{#2}
\let\VANthebibliography\thebibliography
\def\thebibliography{\DeclareRobustCommand{\VAN}[3]{##3}\VANthebibliography}

\usepackage{booktabs}
\usepackage{amsmath}
\usepackage{subfig}
\usepackage{orcidlink}
\usepackage{bm}
\usepackage{multirow}
\usepackage{threeparttable}
\usepackage{array}  

\title{Probing the Limits of Habitability: A Catalog of Rocky Exoplanets in the Habitable Zone}

\author[Bohl et al.]{
    {Abigail Bohl \orcidlink{0009-0006-7156-3152} $^{1}$\thanks{This author contributed equally to this work.}}
    {Lucas Lawrence \orcidlink{0009-0005-6827-6382} $^{1}$\footnotemark[1]}
    {Gillis Lowry \orcidlink{0009-0007-1562-2944} $^{1,2}$\footnotemark[1]}
    {Lisa Kaltenegger \orcidlink{0000-0002-0436-1802} $^{2}$ }
\\
$^{1}$Department of Astronomy and Cornell Center for Astrophysics and Planetary Science, Cornell University, Ithaca, NY, 14853, USA\\
$^{2}$Carl Sagan Institute, Cornell University, 302 Space Sciences Building, Ithaca, NY 14853, USA}

\date{Accepted XXX. Received YYY; in original form ZZZ}

\pubyear{\the\year{}}

\begin{document}
\label{firstpage}
\pagerange{\pageref{firstpage}--\pageref{lastpage}}
\maketitle


\begin{abstract}
While most of the 6000 discovered exoplanets are highly unlike the Earth, the first rocky worlds in the Habitable Zone (HZ) provide intriguing targets for the search for life in the cosmos. As detections increase, it is critical to test the empirical HZ as well as its limits using known exoplanets. However, there is not yet a list of rocky worlds that observers can use to test the limits of surface habitability.

We analysed data from {\it Gaia} DR3 and the NASA Exoplanet Archive (NEA) of all known exoplanets, identifying future targets to test limits of habitability through i) orbits near the edges of the HZ, ii) similar irradiation environments to modern Earth, and iii) large eccentricities. We prioritize targets for transmission observations, light curve measurements, and direct imaging, identify the oldest HZ rocky worlds based on the NEA and complementary literature data, and provide theoretical limits for the empirical HZ and a 3D-HZ for each system.

Our analysis shows 45 rocky worlds in the empirical HZ and 24 in a narrower 3D-HZ. For context, we compare their demographics to those of the full catalog of exoplanets in the NEA.

The resulting list of rocky exoplanet targets in the HZ will allow observers to shape and optimize search strategies with space- and ground-based telescopes—such as the James Webb Space Telescope (JWST), Extremely Large Telescope (ELT), Habitable Worlds Observatory (HWO), and LIFE—and design new observing strategies and instruments to explore these worlds, addressing the question of the limits of {\it exoplanet surface habitability.}

\end{abstract}


\section{Introduction} \label{sec:intro}

Several successful ground- and space-based searches have increased the number of known exoplanets to over 6000 \citep{nasaexoplanetarchivedata}.
%
However, an unexplored aspect of these discoveries is that the growing number of exoplanets allows observers to build a target list of planets that can probe the limits of the Habitable Zone (HZ) empirically. 

The HZ is defined as the orbital range around one or multiple stars at which liquid water could be stable on a rocky planet’s surface \cite[e.g.,][]{Hart1979,Kasting1993,Abe2011,kopparapu2013,Cullum2014, ramirez2014,ramirez2016,Cullum2016,Ramirez2018}, facilitating detection of possible atmospheric biosignatures \cite[e.g.,][]{Kaltenegger2017, Fujii2018, Schwieterman2018, Lichtenberg2024}.  
%
The HZ has been modelled in 1D to 3D for a wide range of stars, with insightful results probing different aspects of planet characterization: 1D models can explore a wide parameter space while 3D GCM models can provide insightful views into the influence of dynamics, relative humidity, and initial cloud feedback, but require more assumptions in topography and rotation rate \cite[e.g.,][]{Kasting1993, Abe2011, kopparapu2013, Leconte2013, Leconte2015, kopparapu2014, kopparapu2016, wolftoon2014, ramirez2014, barnes2015, turbet2017, Yang2023}. Thus, both 1D and 3D models provide important insights into the nature of exoplanets and work in combination to explore new worlds.
While the exploration of models is ongoing, we use the limits of the Empirical HZ defined by \cite{Kasting1993} and \cite{kopparapu2013}, extended to hot stars by \cite{Ramirez2018}. For comparison, we provide an additional, narrower 3D model inner limit defined by \cite{Leconte2013} for Earth-like planets and parametrized into a polynomial representation for different stellar host stars by \cite{Ramirez2018}. Thus, readers can create specific target lists using the values provided (see discussion). 
%

To identify potentially rocky exoplanets, we adopt a maximum radius of 2 $R_\oplus$ \cite[see also][]{hill2023}. Note that the discussion on the proposed radii value that indicates rocky composition is ongoing, with limits between 1.6 and 1.96\,R$_{\oplus}$ \cite[e.g.,][]{Rogers2015, Wolfgang2016, Lehmer2017, Kaltenegger2017, Luque2022, Muller2024}. For planets without measured radii, we adopt a maximum mass of $5\,M_{\oplus}$ in our analysis \cite[see also][]{hill2023}.

Several catalogs \cite[e.g.,][]{Kaltenegger2011,kane2016,tess2018,hill2023} have identified exoplanets in the HZ for earlier epochs, and interesting recent research has specifically focused on assessing dynamical viability \citep{Kane2024}, the properties of the 164 target stars for HWO \citep{Harada2024,mamajek2023}, and threats to HZ exoplanetary systems within 10\,pc \citep{Pyne2024}, as well as exploring the effects of stellar magnetism on the habitability of exoplanets \citep{Atkinson2024} and the limitations of potential UV surface flux \citep{li2024ultravioletphotometryhabitablezones}. 
However, the list of potentially habitable planets has grown since those publications, and new stellar data for host stars are available from the {\it Gaia} DR3 release. In addition, discussions of the influence of the chosen stellar parameters (DR3 compared to NEA),  measurement uncertainty on the list of potentially habitable targets \cite[e.g.,][]{Kaltenegger2011}, and critical analysis of which targets could test our understanding of the edges of the HZ have been missing to prioritize further observations of potential habitable worlds.
%

In this paper, we present a target list of 45 exoplanets in the empirical HZ (27 transiting) and 24 in a narrower 3D HZ (15 transiting). We identify the best exoplanets that can test the limits of surface habitability, orbiting close to the theoretical inner and outer limits of the empirical HZ (see detailed discussion on HZ choices in methods).  We also identify exoplanets that can test how eccentricity influences habitability, exoplanets that receive similar flux to modern Earth, and priority targets for transmission observations, lightcurve measurements, and direct imaging.

\section{Methods} \label{sec:Data}
To identify potentially rocky planets in the HZ, 
we downloaded and analyzed the default values from the NASA Exoplanet Archive (NEA) (date: December 20 2025) \citep{nasaexoplanetarchivedata} for all confirmed 4524 exoplanet systems and 6065 unique exoplanets. 

When available, we updated the stellar data from the NEA (temperature and radius) with results from {\it Gaia} DR3 \citep{gaiadr3datadescription} for stars with re-normalized unit weight error (RUWE) $\leq 1.4$ by cross-referencing {\it Gaia} DR3 data with DR3 designations from the NASA Exoplanet Archive (see discussion). Updates to the host star radius with {\it Gaia} DR3 translate into a corresponding reduction or increase in radius of transiting exoplanets (see discussion).
%
We removed 345 host stars without effective temperature ($\bm{T}_\text{eff}$) in {\it Gaia} DR3 or the NEA from our analysis, because the HZ limits are sensitive to $\bm{T}_\text{eff}$ \cite[e.g.,][]{Kasting1993,Kaltenegger2011}. 
%
We consistently calculated stellar luminosity ($\bm{L}_\text{star}$) from stellar radius ($\bm{R}_\text{star}$) and $\bm{T}_\text{eff}$, and any missing semi-major axis values from stellar mass ($\bm{M}_\text{star}$) and planet orbital period ($\bm{P}_\text{orb}$). This selection provides a sample of 4078 host stars with $\bm{T}_\text{eff}$ between 2566\,K and 7184\,K. 

Planets in multiple-star systems are flagged in the NEA and also in our target list (available at \cite{bohl_zenodo}). None of the final targets of rocky planets in the HZ orbit multiple stars. When comparing with the demographics of the whole exoplanet sample, we did not recalculate the HZ for multiple hosts but used the associated stellar parameters in the NEA, which typically correspond to only one star in the system, even if the exoplanet is flagged as orbiting multiple stars. We undertook an in-depth analysis only for the final sample of rocky planets in the HZ, which do not include multiple hosts according to NEA. The limits of the HZ for exoplanets in multiple-star systems can be assessed following, e.g., \cite{KalteneggerHaghighipour2013}, \cite{HaghighipourKaltenegger2013}, and \cite{KaneHinkel2013}, if observers are interested in these candidates.

%
A subset of planets in the NEA is flagged as eccentric. To include eccentricity, we calculate the time-averaged flux values to assess the averaged incident irradiation for those planets following \cite{Bolmont2016}. If the NEA only provides upper limits on the eccentricity, we do not include them in our calculation but set the eccentricity to zero to no overestimate the eccentricity.

%
The flux boundaries of the HZ can be expressed as a polynomial fit depending on stellar temperature \cite[e.g.,][]{Kasting1993}. 
The HZ limits we focus on in our analysis are based on i) observations in our own Solar System (empirical HZ) \cite[e.g.,][]{Kasting1993,kopparapu2013,Ramirez2018, Kaltenegger2017} and ii) a representative 3D-GCM model \citep{Leconte2013, ramirez2016}. We calculate the empirical HZ flux limits as well as the 3D inner HZ flux limit and compare these limits with the stellar flux the planet receives (see Table \ref{tab:tableRockyHZ}). 

The empirical HZ limits we chose for our analysis are based on solar irradiation when neither a young Venus (Recent Venus, RV) nor a young Mars (Early Mars, EM) had liquid surface water \citep{Kasting1993, kopparapu2013}. The RV limit corresponds to a flux equivalent of 1.76 present-day solar irradiance at Earth’s orbit ($S_0$), and the EM limit to about 0.32\,$S_0$. These flux values correspond to 0.75\,AU and 1.77\,AU respectively in our Solar System, which excludes present-day Venus but includes present-day Mars.
The 3D global climate model we chose for our analysis \citep{Leconte2013} has been specifically developed to quantify the climate response of Earth-like planets to increased insolation in hot and extremely moist atmospheres, and has a polynomial representation for different stellar host stars \citep{ramirez2016}.

The two limits at the outer edge of the HZ are nearly the same, and thus, we only show the empirical HZ limit for the outer edge.
%
However, note that the inner limit of the HZ is debated \citep{Kasting2014}. It could correspond to lower solar flux than the empirical RV limit, but cannot be assessed directly because of Venus' young surface. Thus, young Venus could have lost its water much earlier, or it could have never had liquid surface water \citep{turbet2021}. An idealized 3D land-planet model for dry "Dune" planets, which could be mostly desert with water-rich areas near their poles, shows that such planets could be habitable to 0.77 AU from the Sun, almost at the inner edge of the empirical HZ, because of a much weaker positive water vapor feedback \citep{Abe2011}. Note that some 1D models suggested that the inner edge of the HZ might be even closer to the Sun, as close as 0.38\,AU \citep{Zsom2013}, but that result remains highly controversial \citep{Kasting2014}. Regarding the largest possible distance of the inner HZ limit from the Sun, a conservative theoretical 1D Runaway Greenhouse would put the inner limit at 0.99\,AU for our Solar System, but is arguably too far from the Sun if cloud feedback cools the surface of an Earth-like planet, as expected \cite[e.g.,][]{kopparapu2013, Kaltenegger2017}.
%
The maximum time spent in the HZ we show in Table \ref{tab:tableRockyHZ} is the percentage of the orbital period of each planet within the HZ limits, calculated by numerically solving Kepler’s equation. 

Note that we focus on the HZ limits for Earth-like atmospheres that are dominated by N$_2$-H$_2$O-CO$_2$ here. Additional greenhouse gases, such as significant amounts of molecular hydrogen (H$_2$) \cite[e.g.,][]{Stevenson1999, PierrehumbertGaidos2011,ramirez2017}, CH$_4$ \citep{Ramirez2018}, and ZnS \citep{Kopparapu2018} can extend or curtail the outer edge of the HZ, an effect that is not considered in this study, where we focus on atmospheres similar to Earth's.
Differences in surface pressure and gravity for planets between 0.5 to 5\,M$_{\oplus}$, the mass limit we chose in our analysis, only change the HZ limits by up to 4\% \citep{kopparapu2014} and this effect has, therefore, not been included in our analysis.

To explore which worlds might be at a similar or more advanced stage of evolution compared to life on Earth, we first compile data from the NASA Exoplanet Archive, where 2946 of 6065 exoplanets have age estimates for their host star. We then supplement the missing values through a literature search for the final set of rocky HZ exoplanet hosts. 
Two methods are predominantly used to estimate stellar ages: gyrochronological and isochronal age estimates, based on the slowing spin rate or the star’s properties compared to stellar models, respectively. 
%
Due to cool M stars' extremely long lifespans of 100\,Gyr or more, characteristics such as luminosity, radius, gravitational acceleration, and temperature do not change strongly after entering the main sequence, so isochronal methods are highly inaccurate. Thus, for M and K dwarfs' age estimates, we adopt gyrochronological estimates over isochronal ones; for G dwarfs, we use available isochronal estimates over gyrochronological ones. However, we note that gyrochronological ages may also be an underestimate, as the potential gravitational influence of exoplanets on their host stars can transfer angular momentum, causing these stars to possibly spin more quickly than a planet-less star of the same age \citep{Maxted2015}. 
For cases with multiple age estimates, we report the value from the most recent papers using our chosen methodology and cite our sources.


To assess exoplanets' potential for observations, we calculated the transmission spectroscopy metric (TSM) following \cite{Kempton2018} to guide transit observations, and the maximum apparent angular separation ($\theta$) in combination with the star-planet contrast ratio as guidelines for target selection for direct imaging and secondary eclipse measurements. 
%
If these calculations require missing planetary mass or radius, we used a mass-radius relation of $M~=~0.993\,R^{3.7}$ \citep{Zeng2016} below $2\,R_\oplus$ or $5\,M_\oplus$, and an empirical mass-radius relation for larger planets \citep{Chen2017,Louie2018}, which we use for comparing the subset of rocky HZ planets with the overall exoplanet demographics (TSM values, ($\theta$)s and star-planet contrast ratios). 
For all TSM calculations we used a scale factor of 0.167 for rocky planets, and scale factors binned by planet radii for all non-rocky planets (following \citep{Kempton2018}). 
To calculate the maximum apparent angular separation of the planets, we use the nominal values of the semi-major axis of the planet and the nominal distance to the planetary system, but do not include the effect of eccentricity.
To calculate the star-to-planet contrast ratio, we use the nominal values for the star and planet and assume an Earth-analog bond albedo of 0.3 for all exoplanets, which we assume can be approximated as Lambertian spheres. Note that the contrast ratio can thus easily be scaled for any other albedo values.


Our analysis focuses on the target list of potentially habitable planets in the HZ, but we also discuss which systems would benefit from further observations, because the uncertainties on the stellar and planetary measurements given in the NEA and DR3 allow for the possible inclusion of these exoplanets in the HZ, even though with nominal values they are not placed in the HZ. We use the term “nominal” here to refer to the value listed in the curated NEA table, as well as the curated {\it Gaia} value if the stellar parameters are updated through DR3.
Note that we include this discussion to motivate future observation of these candidates to constrain the uncertainties further, thereby excluding or including them in the list of targets for rocky planets in the HZ. To estimate the maximum error bars on the incident stellar flux at the planet's orbit, we explore the full uncertainty range of the measurements to calculate the smallest and largest possible stellar flux received by the planets: e.g., the smallest possible incident flux value is given for the minimum stellar temperature combined with the minimum stellar radius and the maximum orbital distance. These values are reported as Min $S_0$ for the minimum possible flux and as Max $S_0$ for the maximum possible flux. This simplified approach shows that further observations for these stars could constrain these uncertainties further and could exclude these potential candidates.

\section{Results} 

We analysed the data from the NASA Exoplanet Archive (NEA) of all known exoplanets, identifying \ref{threeone}) targets to prioritize for transmission observations, light curve measurements, and direct imaging, \ref{threetwo}) those that allow exploration of the limits of habitability, and those that provide similar irradiation environments to modern Earth, \ref{threethree}) those that can probe the effect of eccentricity on habitability, and \ref{threefour}) those that are potentially evolved rocky worlds.
We also provide the theoretical limits for the empirical HZ and a 3D-HZ for each system, and identify the oldest HZ rocky worlds based on data from the NEA and complementary literature data.

We identify a total of 290 exoplanets in the empirical HZ and a subset of 45 rocky worlds in the empirical HZ, which provides our target list, and 24 in a narrower 3D-HZ (see Table \ref{tab:tableRockyHZ}). 
27 rocky exoplanets in the empirical HZ transit, compared to 15 rocky exoplanets in the narrower 3D HZ.
%

%
\begin{figure} 
    \centering
    \includegraphics[width=1\linewidth]{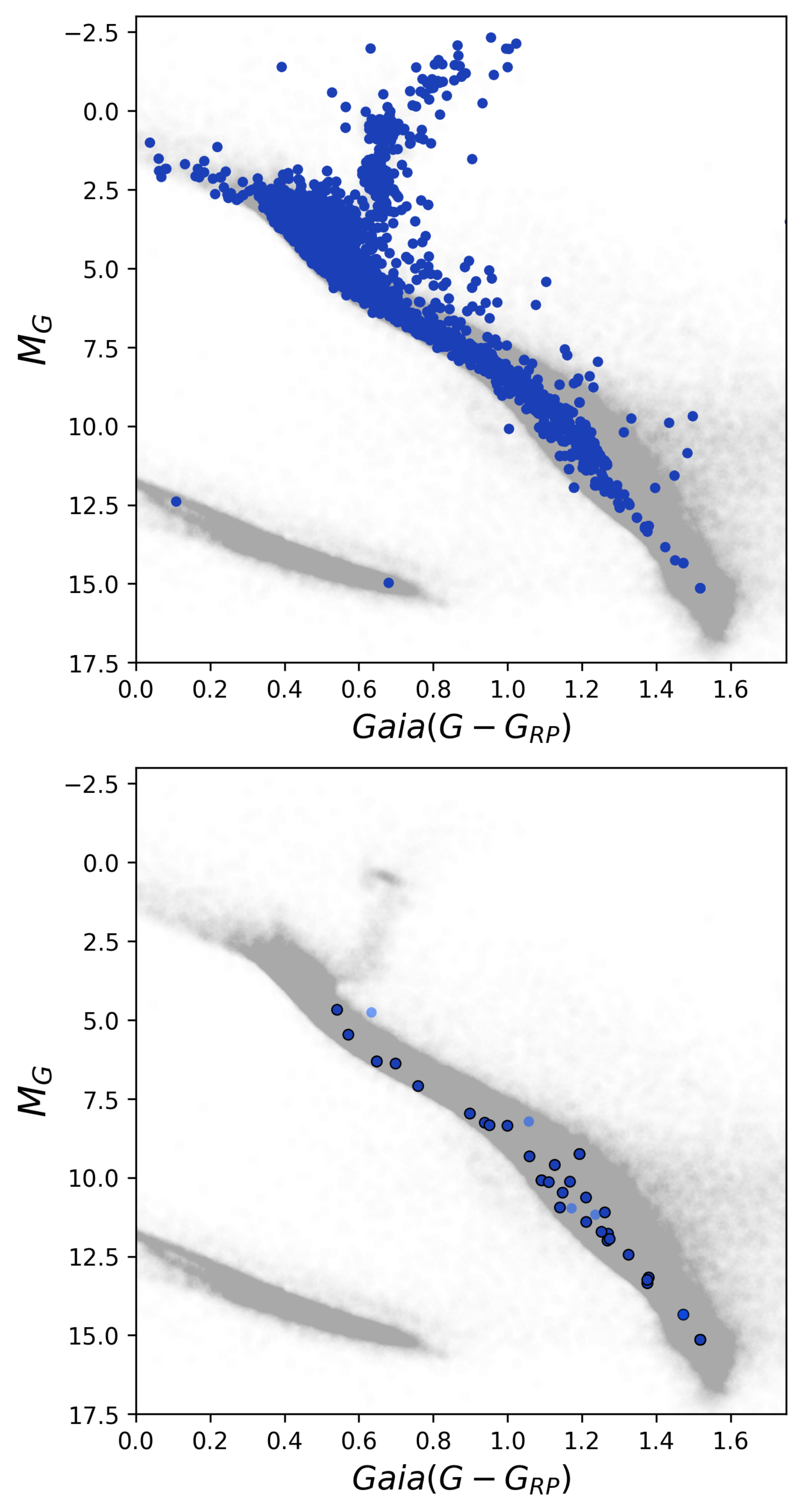}
    \caption{Color-magnitude diagram of all exoplanet host stars (top) and rocky HZ exoplanet host stars (bottom) compared to all stars with {\it Gaia} DR3 data within 100\,pc.}
    \label{fig:NasavsGaiaallstars}
\end{figure}

For context, we further compare the demographics of the rocky exoplanet sample in the HZ to those of the full catalog of exoplanets in the NEA. Figure \ref{fig:NasavsGaiaallstars} shows a color-magnitude diagram for all exoplanet host stars compared to the {\it Gaia} DR3 star sample within 100\,pc, as well as the subset of host stars of rocky exoplanets in the empirical HZ. 
While evolved stars are known to host exoplanets \cite[][e.g.,]{jones2014,Vanderburg2020,Dollinger2021}, no rocky HZ planet has yet been found orbiting evolved stars, although such planets would provide compelling targets for biosignature searches (\cite{kaltenegger2020WD}). The star slightly to the right of the main sequence in the rocky host star panel of Figure \ref{fig:NasavsGaiaallstars} is K2-288 B b, a multiple-star system. 

\begin{figure}
    \centering
            \includegraphics[width=\linewidth]{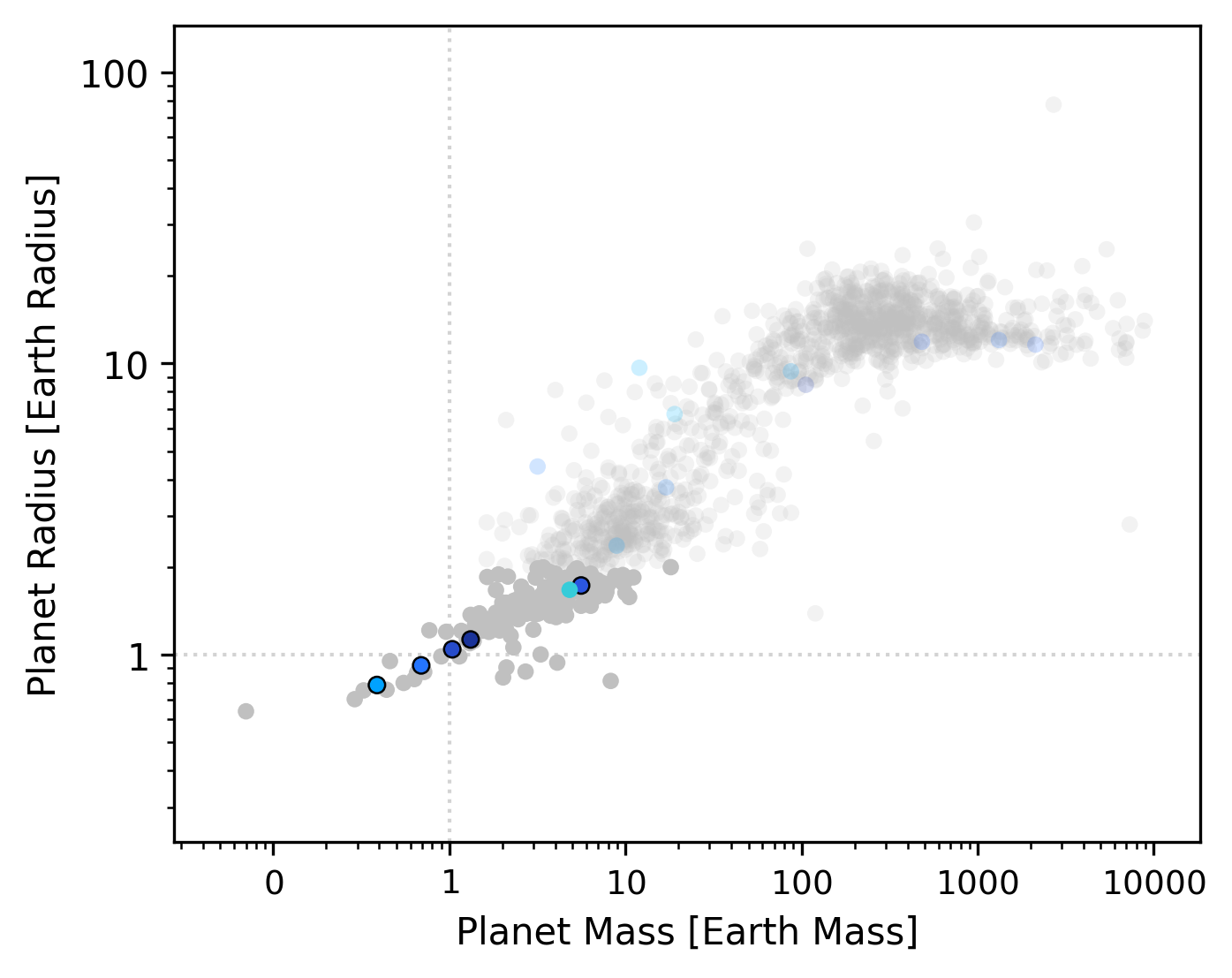}%
            \label{subfig:a}%
        \hfill
            \includegraphics[width=\linewidth]{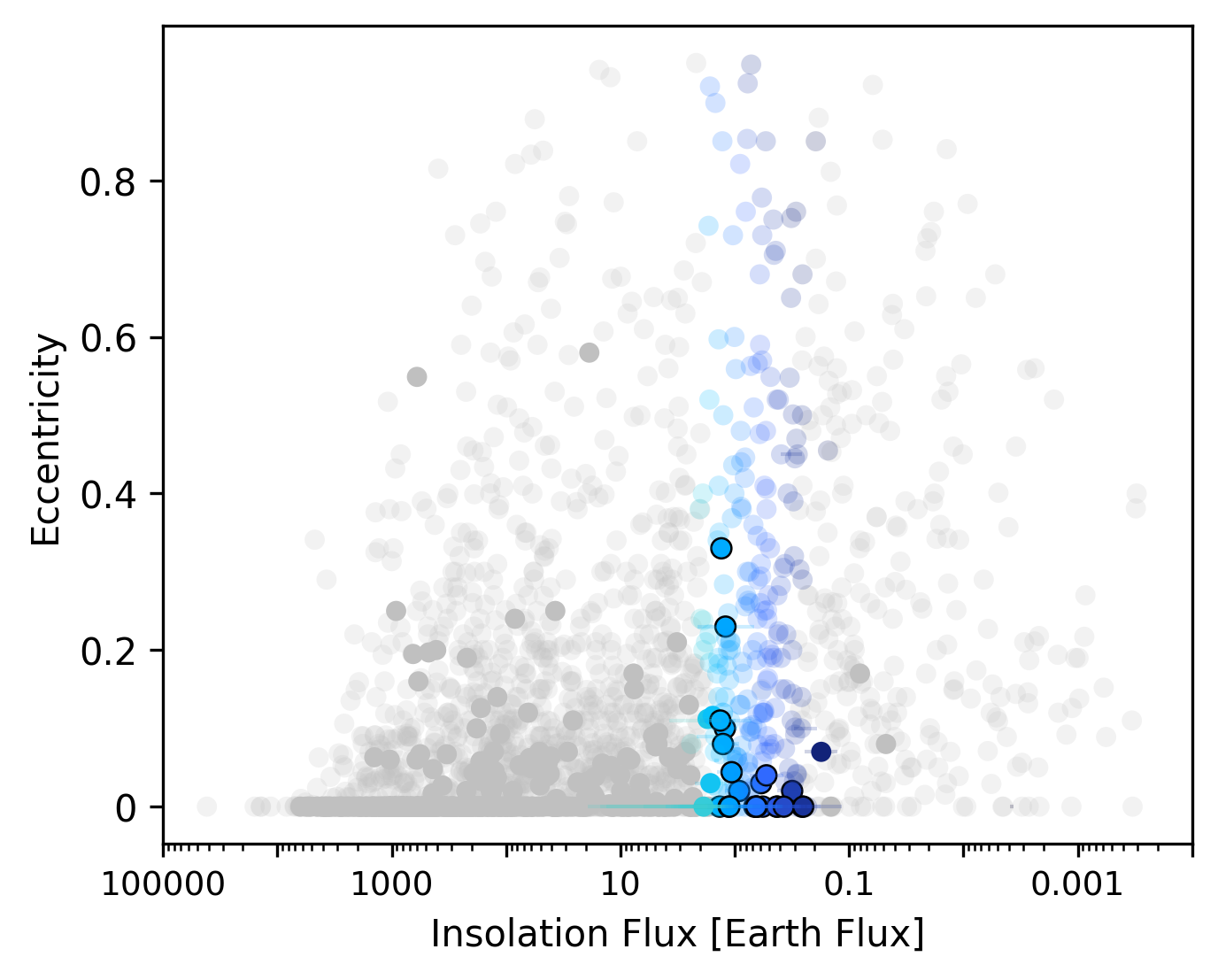}%
            \label{subfig:b}%
        \\
            \includegraphics[width=.5\linewidth]{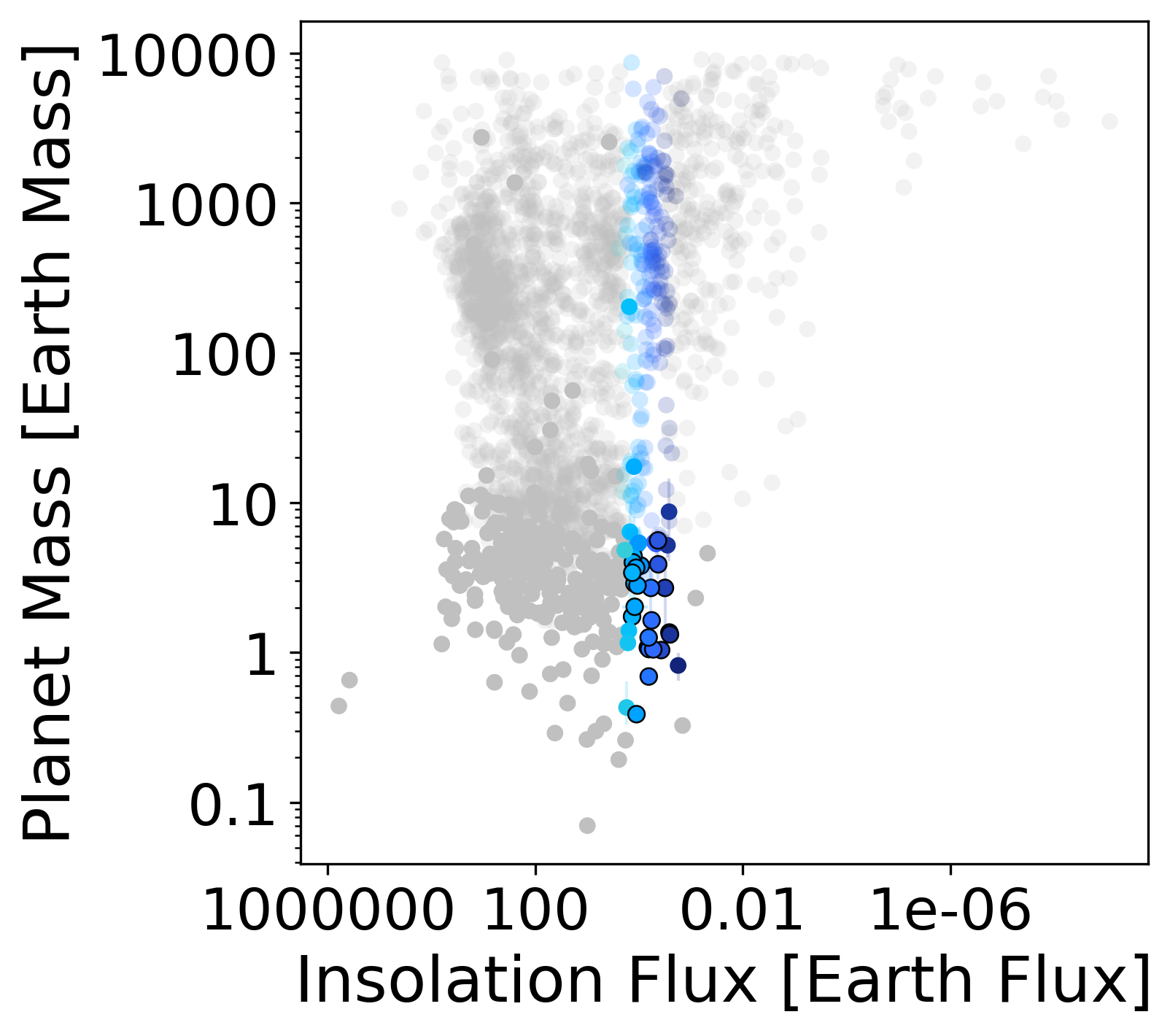}%
            \label{subfig:c}%
        \hfill
            \includegraphics[width=.5\linewidth]{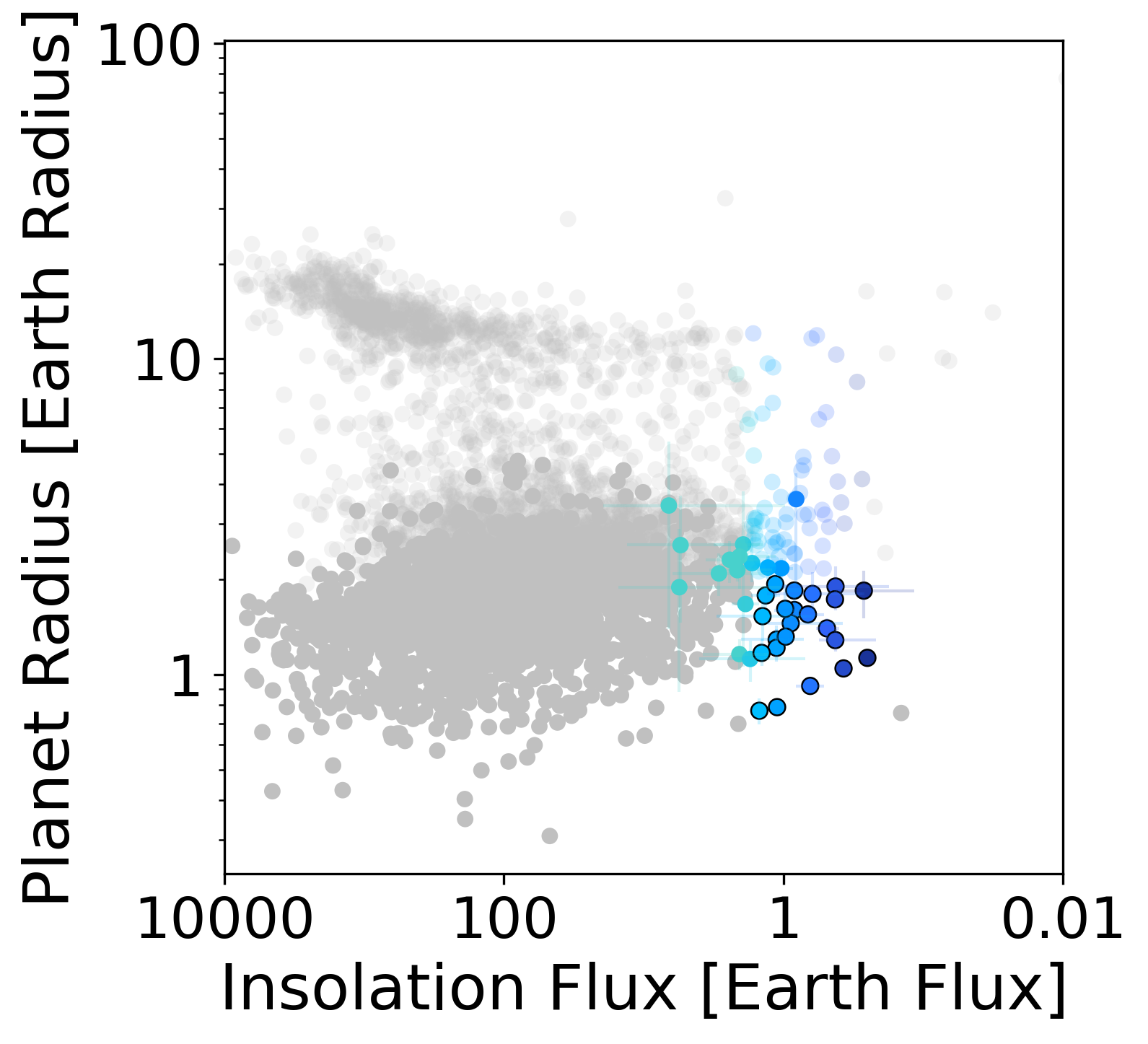}%
            \label{subfig:d}%
        \hfill
 
    \caption{Comparison of the subset of rocky HZ exoplanets (blue) to all known exoplanets (grey): mass vs. radius (top), eccentricity vs. incident flux (middle), minimum mass vs. incident flux (bottom left), and radius vs. incident flux (right). 
    HZ planets are shown in blue, planets in the 3D HZ limits in dark blue, and all other known exoplanets in grey. Rocky planets are solid-color, while non-rocky planets are semi-transparent. Planets with only upper mass limits are not shown in the mass-radius plot. 
    }
    \label{fig:demographics}
\end{figure}

To put the rocky HZ exoplanets in context, we show their parameters (colored) compared to all known exoplanets (grey) in Figure \ref{fig:demographics}. The subset of exoplanets in the empirical HZ is shown in blue, and exoplanets in the 3D HZ are shown in dark blue. Rocky HZ planets in our sample are defined with $\leq2\,R_\oplus$, or $\leq5\,M_\oplus$ for non-transiting planets, and shown as solid colors. 
%

The radius-mass plot for all exoplanets with measured mass and radius values is shown in the top left panel of Figure \ref{fig:demographics}, and excludes planets with only an upper limit on their mass. It identifies five rocky HZ exoplanets with both measured mass and radius: LHS~1140~b orbiting an M4.5 dwarf star at 15\,pc, and TRAPPIST\nobreakdash-1d, \nobreakdash-1e, \nobreakdash-1f, and \nobreakdash-1g orbiting an M8 dwarf star at 12\,pc. 
These five exoplanets have been the focus of JWST observations \cite[e.g.,][]{Lim2023,Benneke2023,Greene2023,Cadieux2024}, which provided first insights into the atmospheric composition of rocky planets in the HZ of these systems.
%
TRAPPIST\nobreakdash-1d, \nobreakdash-1e, \nobreakdash-1f, and \nobreakdash-1g have nominal mass and radii values of $0.79\,R_\oplus$/$0.39\,M_\oplus$, $0.92\,R_\oplus$/$0.69\,M_\oplus$, $1.05\,R_\oplus$/$1.04\,M_\oplus$, and $1.13\,R_\oplus$/$1.32\,M_\oplus$, respectively. All four exoplanets have slightly lower nominal density than Earth: 0.79\,$\rho_\oplus$, 0.89\,$\rho_\oplus$, 0.91\,$\rho_\oplus$, and 0.92\,$\rho_\oplus$, respectively.
%
%
LHS~1140~b has a nominal radius of $1.73\,R_\oplus$ and a nominal mass of $5.6\,M_\oplus$, resulting in a slightly higher density than Earth's: $1.08\,\rho_\oplus$. 

The bottom panels of Figure \ref{fig:demographics} show the radius distribution as well as the mass distribution of known exoplanets, showing patterns of planet demographics mixed with detection biases, like the broad transition zone between the mass of a mini-Neptune and a gas giant like Jupiter, the smaller number of exoplanets detected at larger orbital periods and lower masses and radius. Rocky exoplanets in the HZ are found in the region of lower stellar incident irradiation in both plots. Note that, especially, the distribution of rocky exoplanets in the HZ is still strongly influenced by the limited number of observations, and thus only shows a first indication of the underlying distribution.

\begin{figure*}
    \centering
    \includegraphics[width=1\linewidth]{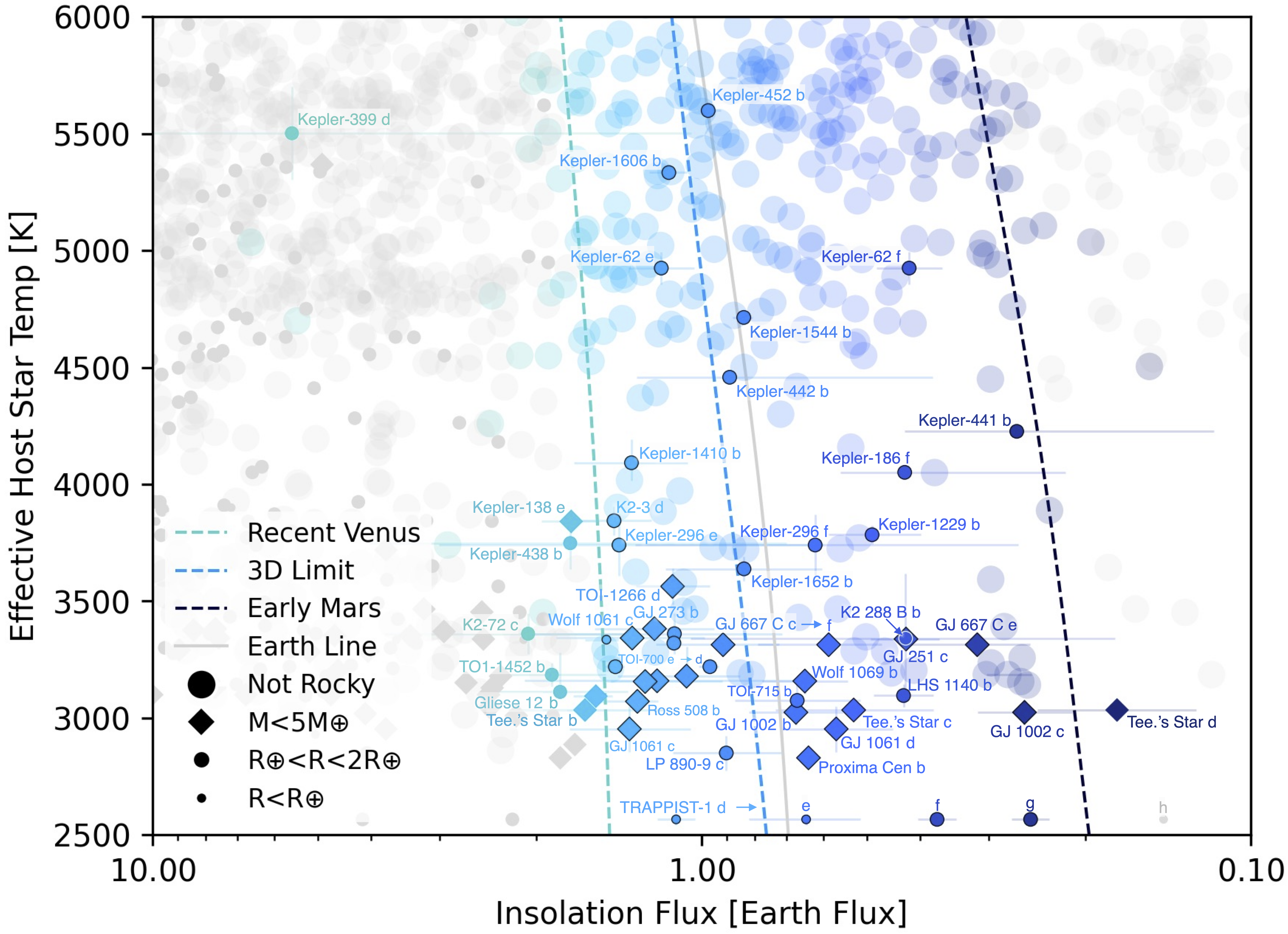}
    \caption{All known exoplanets shown in terms of their host star's temperature and the incident stellar flux they receive. Transiting rocky exoplanets are shown as circles, and planets where only minimum mass is known as diamonds. Exoplanets in the HZ are shown in blue, with exoplanets  $\leq 2\,R_{\oplus}$ and $\leq 5\,M_{\oplus}$ shown as smaller symbols in solid colors. Empirical HZ limits and the 3D inner HZ limit are shown as dashed lines. A solid light grey line indicates similar irradiation to modern Earth for different stellar temperatures.}
    \label{fig:temp-flux}
\end{figure*}

Figure \ref{fig:temp-flux} shows all known exoplanets that receive stellar flux between 0.1 and 10\,$S_0$. The empirical HZ limits and a 3D inner HZ limit are shown as dashed lines. A light grey line indicates similar irradiation to modern Earth for different stellar types. Transiting rocky planets are plotted as circles, while non-transiting planets where only minimum masses are known are plotted as diamonds. Maximum uncertainties on a planet's incident stellar irradiation are shown as error bars as discussed in methods.

The Catalog of Habitable Zone Rocky Planets (see Table \ref{tab:tableRockyHZ}, full table available at \cite{bohl_zenodo}) lists planet data (name, radius and corresponding uncertainty, and mass and corresponding uncertainty), incident stellar flux (average, min, and max flux), flux at two inner limits of the HZ (empirical Recent Venus (RV) and a 3D model (3D) ), and an outer empirical Early Mars (EM) limit. All flux is provided in units of modern Earth flux ($\bm{S_0}$). The maximum and minimum possible stellar flux reaching each planet is calculated based on the measurement uncertainties in stellar temperature ($\bm{T}_\text{eff}$), semi-major axis, and nominal orbital eccentricity. Apparent angular separation ($\bm \theta$), planet-star contrast ratio, a stellar age estimate, distance to the system in pc ($\bm d$), and ($\bm{T}_\text{eff}$) are also given. 

The sample of rocky exoplanets in the HZ in Table \ref{tab:tableRockyHZ} is sorted first by descending TSM values for those that transit (above dotted row), and by angular separation for planets with contrast ratio above $10^{-8}$, for direct imaging for those that do not transit (below dotted row). Note that while the apparent angular separation from the host star is one important metric for direct imaging feasibility, the contrast ratio also is important to assess detectability, with both requirements varying by instruments, and target selection can easily be adjusted using the full table that provides among others, contrast ratio, apparent angular separation, and J and Ks 2Mass magnitudes of the star.

Note that the CSV file (available at \cite{bohl_zenodo}) has a much wider range of parameters, among them J and Ks magnitude, multiplicity flags, and references to the mass, radius, and age values. The host stars of the rocky exoplanets in the empirical HZ have apparent magnitudes between 5.36 to 13.88 in the infrared J 2MASS band, 4.38 to 13.40 in the near-infrared Ks 2MASS band, and 9.84 to 18.00 in the visual V Johnson band.

\begin{table*}
    \centering 
    \begin{tabular}{lclccccccrcclcc} 
        \toprule
        Planet Name& Radius & Mass & {Flux} & {Min} & {Max} & {RV} & {3D} & {EM} & {TSM} & {$\theta$} & {Contrast} & {Age} & {d} & {$\bm{T}_\text{eff}$}\\
        & {R$_\oplus$} & {M$_\oplus$} & {$S_0$} & {$S_0$} & {$S_0$} & {S$_0$} & {$S_0$} & {$S_0$} & {} & {mas} & {} & {Gyr} & {pc} & {K}\\
        \midrule
        TRAPPIST-1 d & 0.79$^{\pm0.01}$ & 0.39$^{\pm0.01}$ & 1.11 & 1.03 & 1.21 & 1.47 & 0.77 & 0.20 & 22.75 & 1.78 & 1.71e-7 & 7.6$^{\pm2.2}$& 12.47 & 2566$^{\pm26}$\\
        TRAPPIST-1 e & 0.92$^{\pm0.01}$ & 0.69$^{\pm0.02}$ & 0.65 & 0.51 & 0.82 & 1.47 & 0.77 & 0.20 & 17.71 & 2.35 & 1.35e-7 & 7.6$^{\pm2.2}$ & 12.47 & 2566$^{\pm26}$\\
        TRAPPIST-1 f & 1.05$^{\pm0.01}$ & 1.04$^{\pm0.03}$ & 0.37 & 0.34 & 0.40 & 1.47 & 0.77 & 0.20 & 15.07 & 3.09 & 1.00e-7 & 7.6$^{\pm2.2}$ & 12.47 & 2566$^{\pm26}$\\
        TRAPPIST-1 g & 1.13$^{+0.02}_{-0.01}$ & 1.32$^{\pm0.04}$ & 0.25 & 0.23 & 0.27 & 1.47 & 0.77 & 0.20 & 13.55 & 3.76 & 7.92e-8 & 7.6$^{\pm2.2}$ & 12.47 & 2566$^{\pm26}$\\
        LHS 1140 b & 1.73$^{\pm0.03}$ & 5.6$^{\pm0.19}$ & 0.43 & 0.38 & 0.49 & 1.50 & 0.81 & 0.21 & 8.93 & 6.31 & 4.56e-8 & 7.84$^{\pm3.79}$ & 14.99 & 3096$^{\pm48}$\\
        $\cdots$ & $\cdots$ & $\cdots$ & $\cdots$ & $\cdots$ & $\cdots$ & $\cdots$ & $\cdots$ & $\cdots$ & $\cdots$ & $\cdots$ & $\cdots$ & $\cdots$ & $\cdots$ & $\cdots$\\
        Proxima Cen b & & 1.06$^{\pm0.06}$ & 0.64 & 0.63 & 0.65 & 1.49 & 0.79 &  0.20 & 134.42 & 37.26 & 5.99e-8 & 5.2$^{\pm1.61}$ & 1.30 & 2829.35$^{+0.21}_{-0.36}$\\
        Wolf 1061 c & & 3.41$^{+0.43}_{-0.41}$ & 1.34 & 0.98 & 1.83 & 1.51 & 0.83 & 0.22 & 27.31 & 20.67 & 3.35e-8 & 6.38$^{\pm2.53}$ & 4.31 & 3342$^{\pm49}$ \\
        GJ 667 C c &  & 3.8$^{+1.5}_{-1.2}$ & 0.92 & 0.72 & 1.21 & 1.51 & 0.83 & 0.22 & 11.48 & 17.25 & 1.80e-8 & 2$^{+8}$ & 7.24 & 3313.44$^{+36.06}_{-2.41}$ \\   
        GJ 682 b & & 4.4 $^{+ 3.7 }_{ -2.4 }$& 1.27 & 0.91 & 1.46 & 1.50 & 0.82 & 0.21 & 20.25 & 15.99 & 4.76e-08 & 6.4$^{\pm4.3}$ & 5.01 & 3155.80 $^{+ 1.41 }_{ -0.50 }$\\
        GJ 273 b & & 2.89 $^{+ 0.27 }_{ -0.26 }$ & 1.22 & 0.94 & 1.56 & 1.52 & 0.84 & 0.22 & 33.69 & 15.38 & 2.93e-08 & 10.31 $^{\pm 6.2 }$ & 5.92 $^{\pm 0.02 }$ & 3382 $^{\pm 49.0 }$\\
        \bottomrule
    \end{tabular}
    \begin{tablenotes}
	\small {
	\item{* Indicates controversial planets in the NEA (see discussion)}}
    \end{tablenotes}
    \caption{Sample table of the Catalog of Habitable Zone Rocky Planets. Planets are sorted first by whether their nominal values place them as rocky HZ planets, then by descending TSM values for those that transit (above dotted row), and by apparent angular separation for planets with contrast ratio above $10^-8$ for direct imaging of those that do not transit (below dotted row). 
    Incident flux (Flux) is provided in units of modern Earth flux ($S_0$), along with the minimum and maximum value (Min, Max), two inner limits of the HZ—the empirical Recent Venus (RV) and a 3D model (3D) limit— and the outer empirical Early Mars (EM) limit. The maximum and minimum possible stellar flux reaching each planet is calculated based on measurement uncertainties in stellar temperature ($\bm{T}_\text{eff}$), uncertainties in the semi-major axis, and nominal eccentricity. 
    The full table is available at \citet{bohl_zenodo}.}
    \label{tab:tableRockyHZ}
\end{table*}

\subsection{Best Transiting and Direct Imaging Rocky HZ Exoplanets}\label{threeone}

%
To prioritize exoplanets for follow-up observations, we calculate three metrics that identify the best targets for transmission spectroscopy, light curve measurements, and direct imaging. The transmission spectroscopy metric (TSM) identifies the best targets for transit observations, and the apparent angular separation ($\bm\theta$) in combination with the contrast ratio between the host star and its planet identifies the best targets for direct imaging observations and secondary eclipse and light curve measurements. 

\begin{figure}
    \centering
    \includegraphics[width=1\linewidth]{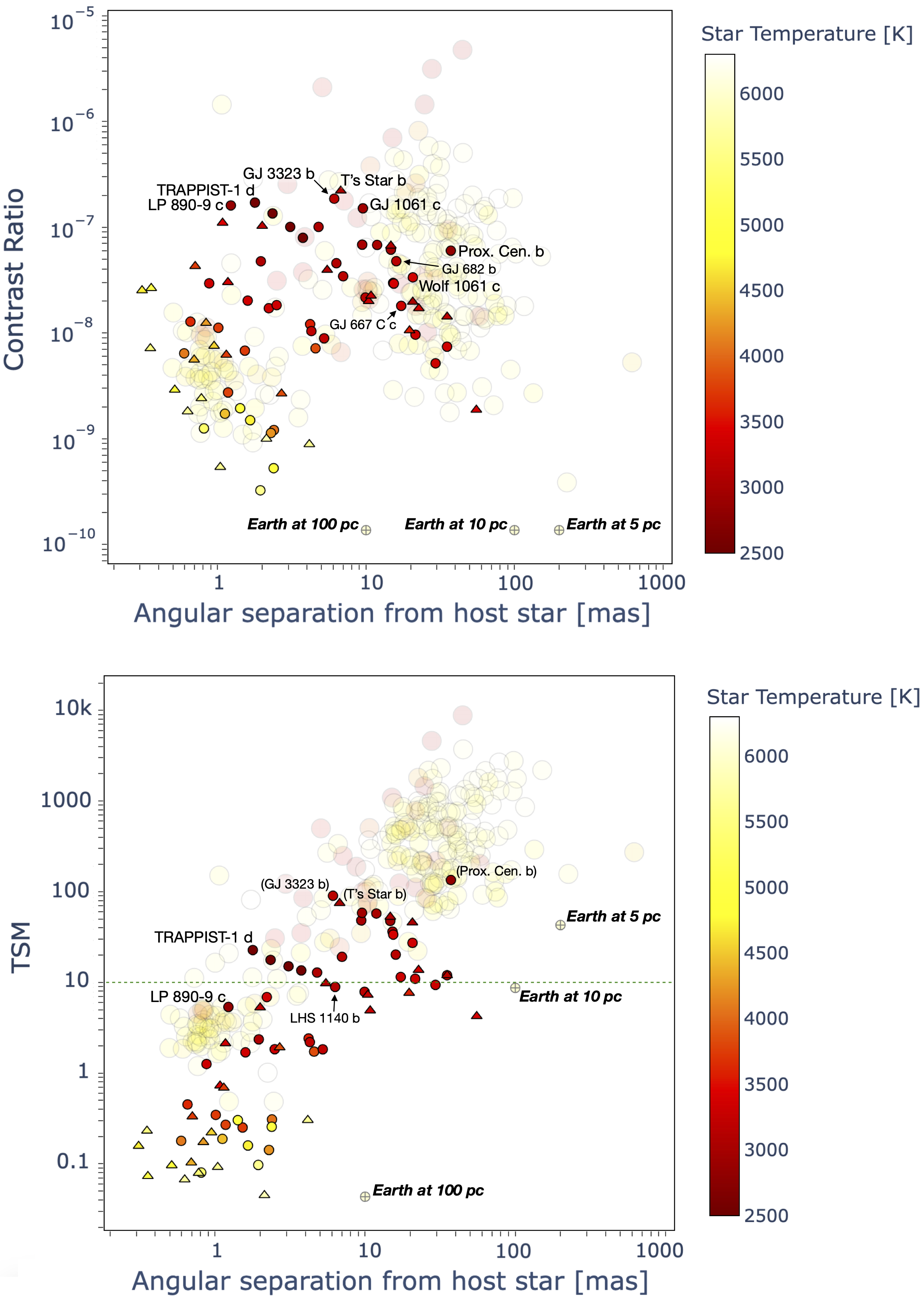}
    \caption{Apparent angular separation of rocky exoplanets in the HZ plotted versus contrast ratio between host star and planet (top) and versus transmission spectroscopy metric (TSM) (bottom). Rocky planets in the HZ with nominal values are plotted as solid-color circles, and rocky planets in the HZ including measurement uncertainties (in irradiation and planet size) are plotted as triangles. Other HZ planets are larger, semi-transparent circles. A TSM value of 10 is shown as a green dashed line.}
    \label{fig:TSM}
\end{figure}

Table \ref{tab:contrast-tsm-ang} shows the five best targets for transiting (left) and five best tragets for non-transiting (right) rocky planets, sorted by contrast ratio (top) and angular resolution (middle), and by TSM (for transiting) and by angular resolution for  contrast ratios larger than 10$^{-8}$ (direct imaging, secondary eclipse, lightcurves) (bottom).
Figure \ref{fig:TSM} shows the $\bm\theta$ of each exoplanet versus its contrast ratio (top) and the TSM value of each planet versus its contrast ratio (bottom), color-coded for host star temperature. Earth's $\bm\theta$, contrast ratio, and TSM value at 5, 10, and 100\,pc are shown for comparison. For consistency, here Earth's temperature and TSM value are calculated from the J band magnitude of 3.67 \citep{Willmer2018} for the Sun. 

\begin{table}
\centering
\begin{tabular}{ll c lc}
\toprule
{Transiting} & {Value} & \vrule & {Direct Imaging} & {Value}\\
\midrule        
\multicolumn{5}{c}{\textbf{Best Rocky HZ Planets Sorted by Contrast}} \\
\midrule

    TRAPPIST\nobreakdash-1~d & 1.7e\nobreakdash-7 &\vrule & GJ~3323~b & 1.9e\nobreakdash-7\\
    LP~890-9~c & 1.6e\nobreakdash-7 & \vrule & GJ~1061~c & 1.5e\nobreakdash-7 \\  
     TRAPPIST\nobreakdash-1~e & 1.4e\nobreakdash-7& \vrule & Ross~508~b & 1.0e\nobreakdash-7 \\
    TRAPPIST\nobreakdash-1~f & 1.0e\nobreakdash-7 & \vrule & GJ~1002~b & 6.8e\nobreakdash-8\\
    TRAPPIST\nobreakdash-1~g & 7.9e\nobreakdash-8 & \vrule & Teegarden's~Star~c & 6.8e\nobreakdash-8\\

\midrule      
\multicolumn{5}{c}{\textbf{Best Rocky HZ Planets Sorted by $\theta$ (mas) }} \\
\midrule
    LHS~1140~b & 6.3 & \vrule & Proxima~Cen~b & 37.3 \\
    TOI\nobreakdash-700~d & 5.2 & \vrule & GJ~251~c & 35.1 \\
    K2\nobreakdash-3~d & 4.6 & \vrule & GJ~667~C~e* & 29.4  \\
    TOI\nobreakdash-700~e & 4.3 &\vrule & GJ~667~C~f* & 21.5 \\
    TRAPPIST\nobreakdash-1~g & 3.8 & \vrule & Wolf~1061~c & 20.7 \\

\midrule 

\multicolumn{2}{c}{\textbf{Best Transits by TSM}} & \multicolumn{1}{c}{\vrule} & \multicolumn{2}{c}{\textbf{Best Direct Image by $\theta$, Contrast}} \\

\midrule
    TRAPPIST\nobreakdash-1~d & 22.8 & \vrule & Proxima~Cen~b & (37.3, 6.0e\nobreakdash-8)\\
    TRAPPIST\nobreakdash-1~e & 17.1 & \vrule & Wolf~1061~c & (20.7, 3.4e\nobreakdash-8) \\ 
    TRAPPIST\nobreakdash-1~f & 15.1 & \vrule & GJ~667~C~c & (17.3, 1.8e\nobreakdash-8) \\
    TRAPPIST\nobreakdash-1~g & 13.6 & \vrule & GJ~682~b & (16.0, 4.8\nobreakdash-8)\\
    LHS~1140~b & 8.9 & \vrule & GJ~273~b & (15.4, 2.9e\nobreakdash-8) \\ 


\bottomrule
\end{tabular}
\begin{tablenotes}
\small {
\item{* Indicates controversial planets in the NEA (see discussion)}}
\end{tablenotes}
\caption {Transiting and non-transiting planet targets from the final rocky HZ planet list ranked by contrast ratio, by angular separation, and by either TSM (for transit observation) or by angular resolution after filtering for contrast ratio above $10^{-8}$ (for direct imaging of non-transiting planets).  
}
\label{tab:contrast-tsm-ang}
\end{table}

\subsection{Best rocky exoplanets to test the limits of the HZ}\label{threetwo}
To explore the limits of surface habitability, Table \ref{tab:tableLimits} shows the best rocky exoplanets near the inner and outer regions of the empirical HZ for transiting (sorted by highest TSM) and for direct imaging (sorted by highest $\bm\theta$ and contrast ratio). In addition to probing the edges of habitability, Table~\ref{tab:tableLimits} shows 10 exoplanets (transiting and non-transiting) that receive stellar insolation similar to Earth's, which can explore whether this provides similar surface conditions.

\begin{table}
    \centering 
    \begin{tabular}{lcccc} 
        \toprule
        Category & Planet Name & {TSM} & {$\theta$} & {Contrast}\\
        {} & {} & {} & {mas} & {}\\
        \midrule
        \textbf{Inner HZ} & & & & \\
        \midrule
        \multirow{3}{*}{Transiting} 
        & K2\nobreakdash-239~d & 6.9 & 2.2 & 1.7e\nobreakdash-8 \\
        & TOI\nobreakdash-700~e & 2.2 & 4.3 & 1.0e\nobreakdash-8 \\ 
        & K2\nobreakdash-3~d & 1.7 & 4.6 & 7.1e\nobreakdash-9 \\
        \midrule
        \multirow{2}{*}{Non-Transiting}
        & Wolf~1061~c & 27.3 & 20.7 & 3.4e\nobreakdash-8 \\ 
        & GJ~1061~c & 58.5 & 9.5 & 1.5e\nobreakdash-7 \\ 
        \midrule
        \textbf{Outer HZ} & & & & \\
        \midrule
        \multirow{3}{*}{Transiting}
        & TRAPPIST\nobreakdash-1~g & 13.6 & 3.8 & 7.9e\nobreakdash-8 \\ 
        & Kepler\nobreakdash-186~f* & 0.3 & 2.4 & 1.2e\nobreakdash-9 \\
        & Kepler\nobreakdash-441~b & 0.1 & 2.3 & 1.1e\nobreakdash-9 \\ 
        \midrule
        \multirow{1}{*}{Non-Transiting}
        & GJ~1002~c & 36.3 & 15.2 & 3.0e\nobreakdash-8 \\ 
        & GJ~667~C~e* & 9.4 & 29.4 & 5.2e\nobreakdash-9 \\ 
        \midrule
        \textbf{Present Earth-like Flux} & & & & \\
        \midrule
        \multirow{5}{*}{Transiting} & 
        TRAPPIST\nobreakdash-1~e & 17.7&  2.3 & 1.3e\nobreakdash-7 \\ 
        & TOI\nobreakdash-715~b & 2.4 & 2.0 & 4.8e\nobreakdash-8 \\ 
        & Kepler\nobreakdash-1652~b & 0.5 &  0.7 & 1.3e\nobreakdash-8\\ 
        & Kepler\nobreakdash-442~b & 0.2 & 1.1 & 1.7e\nobreakdash-9\\ 
        & Kepler\nobreakdash-1544~b & 0.2 & 1.7 & 1.5e\nobreakdash-9 \\
        & Kepler\nobreakdash-452~b* & 0.1&  1.9 & 3.2e\nobreakdash-10 \\ 
        \midrule
        \multirow{3}{*}{Non-Transiting} 
        & Proxima~Cen~b & 134 & 37.3 &  6.0e\nobreakdash-8 \\ 
        & GJ~1061~d & 47.6 & 14.7 & 6.1e\nobreakdash-8 \\ 
        & GJ~1002~b & 44.7 & 9.4 &  6.8e\nobreakdash-8 \\
        & Wolf~1069~b & 19.1 & 7.0 &  3.4e\nobreakdash-8\\
        \bottomrule
    \end{tabular}
    \begin{tablenotes}
	\small {
	\item{* Indicates controversial planets in the NEA (see discussion)}}
    \end{tablenotes}
    \caption{Rocky planets in the HZ with the nearest irradiation values to the HZ boundaries (Inner HZ, Outer HZ) and to modern Earth irradiation (Earth-like flux), sorted by descending TSM for transiting planets and descending angular separation and contrast ratio for non-transiting planets. All exoplanets in the present Earth-like flux category receive stellar fluxes within $\pm15$\% of present Earth line shown in Fig. \ref{fig:temp-flux}. }
    \label{tab:tableLimits}
\end{table}

\subsection{Best rocky exoplanets to test the effects of eccentricity on habitability}\label{threethree}
Eccentric planets could help constrain the effect of eccentricity on habitability \cite[see e.g.,][]{Bolmont2016, Liu2024}. The eccentricity of rocky HZ exoplanets is generally low (see Fig. \ref{fig:demographics}) \citep[e.g.,][]{hill2023}, except for the few targets listed in Table \ref{tab:tableecc}. Observing these planets, especially at different times in their orbit or over a full orbit, could provide insights into how eccentricity influences habitability and how atmospheres react to different stellar irradiation.

In the full table (available at \cite{bohl_zenodo}), we include the calculated time-averaged flux values, which were used to assess whether the planets lie within the HZ limits. We also calculate the time spent in the HZ as the maximum percentage of the orbital period of each planet within the HZ limits, based on measurement uncertainties for semi-major axis, eccentricity, $\bm{T}_\text{eff}$, and $\bm{R}_\text{star}$. 

\begin{table}
    \centering 
    \begin{tabular}{lccccc} 
        \toprule
        Category & Planet Name & Ecc & {TSM} & {$\theta$} & {Contrast}\\
        {} & {} & {} & {} & {mas} & {}\\
        \midrule
       \multirow{1}{*}{Non-transiting} & 
        Ross~508~b & 0.33 & 12.8 & 4.8 & 1.0e\nobreakdash-7 \\ 
        & GJ~3323~b & 0.23 & 90.2 & 6.1 & 1.9e\nobreakdash-7 \\
        & Wolf~1061~c & 0.11 & 27.3 & 4.5 & 3.4e\nobreakdash-8 \\
       \midrule
        \multirow{1}{*}{Transiting} &
        K2-72~e & 0.11 & 1.7 & 1.6 & 2.0e\nobreakdash-8 \\
        \bottomrule
    \end{tabular}
    \caption{Best HZ rocky planets to test the effects of high eccentricity on habitability.}
    \label{tab:tableecc}
\end{table}

\subsection{Best Potentially Evolved Rocky Exoplanets in the HZ}\label{threefour}
Age estimates for exoplanets in the HZ vary significantly and carry large uncertainties. However, even first estimates are useful to compare with Earth's atmosphere and observable spectra at a given time, as these have changed significantly since its formation \citep[see e.g.,][]{Kaltenegger2007, Selsis2007, Rugheimer2015, Rugheimer2018, Kaltenegger2020, Payne2024}.
Although it is unknown what sets the evolution timescale on a rocky exoplanet, comparing tabulated stellar age estimates for hosts of rocky HZ exoplanets to the evolution of Earth allows for a first comparison of their evolutionary stages and timescales.  

In the final list of 45 rocky exoplanets in the empirical HZ, 15 of the 35 host stars were missing stellar age estimates in their NEA default values. Searching the literature, we tabulated missing age estimates for 10 stars, and updated the existing values for 6 stars with more recent values, which use methods that improve estimates for the specific host star's stellar type. Thus, the age estimates of 16 host stars are based on a literature search \citep{Engle2023, Berger2020, Dreizler2020,Maldonado2020,Guillem2012,González-Álvarez2023, Dholakia2024, Gaidos2023, Schlieder2016, Torres2015, Torres2017,Morton2016,Armstrong2016,Jenkins2015,Borucki2013,Delrez2022, Fukui2022, Dransfield2024,Burgasser2017,Kossakowski2023} and differ from the NEA's values. Note that 3 stars only have qualitative tabulated age estimates (GJ~1061, GJ~667~C, Wolf~1069).
5 stars are still missing age estimates (K2-72, Kepler-1649, Kepler-186, Kepler-452, Ross~508).

\begin{table}
    \centering 
    \begin{tabular}{lccccc} 
        \toprule
        Category & Planet Name & Age & {TSM} & {$\theta$} & {Contrast}\\
        {} & {} & {Gyr} & {} & {mas} & {}\\
        \midrule
        \multirow{6}{*}{Transiting} & 
        K2\nobreakdash-239~d & 13.32$^{+10.3}_{-9.5}$ & 6.9 &  2.2 & 1.7e\nobreakdash-8 \\ 
        & LHS~1140~b & 7.84$^{\pm3.8}$ & 6.3 & 6.0 & 4.6e\nobreakdash-8 \\
        & TRAPPIST\nobreakdash-1~d & 7.6$^{\pm2.2}$ & 22.7 &	1.8 & 1.7e\nobreakdash-7 \\
        & TRAPPIST\nobreakdash-1~e & 7.6$^{\pm2.2}$ & 17.7 &	2.3 & 1.3e\nobreakdash-7 \\
        & TRAPPIST\nobreakdash-1~f & 7.6$^{\pm2.2}$ & 15.1 & 3.1 & 1.0e\nobreakdash-7 \\
        & TRAPPIST\nobreakdash-1~g & 7.6$^{\pm2.2}$ & 13.5 & 3.8 & 7.9e\nobreakdash-8 \\
        \midrule
        \multirow{4}{*}{Non-transiting} &
        GJ~273~b & 10.3$^{\pm6.2}$ & 33.7 & 15.4 & 2.9e\nobreakdash-8 \\ 
        & GJ~1002~c & 7.5$^{\pm3.6}$ & 36.3 & 15.2 & 3.0e\nobreakdash-8 \\
        & GJ~1002~b & 7.5$^{\pm3.6}$ & 48.2 & 9.4 & 6.8e\nobreakdash-8 \\
        & GJ~1061~d & 7.0$^{\pm0.5}$ & 47.6 & 14.7 & 6.1e\nobreakdash-8 \\
        & GJ~1061~c & 7.0$^{\pm0.5}$ & 58.5 & 9.5 & 1.5e\nobreakdash-7 \\
        \bottomrule
    \end{tabular}
    \caption{Rocky HZ planet host stars with the oldest estimated nominal ages. Planets with similar ages are ranked by descending TSM for those that transit and descending angular separation for those that do not transit with contrast ratios above $10^{-8}$.}
    \label{tab:age}
\end{table}

\begin{figure}
    \centering
    \includegraphics[width=1\linewidth]{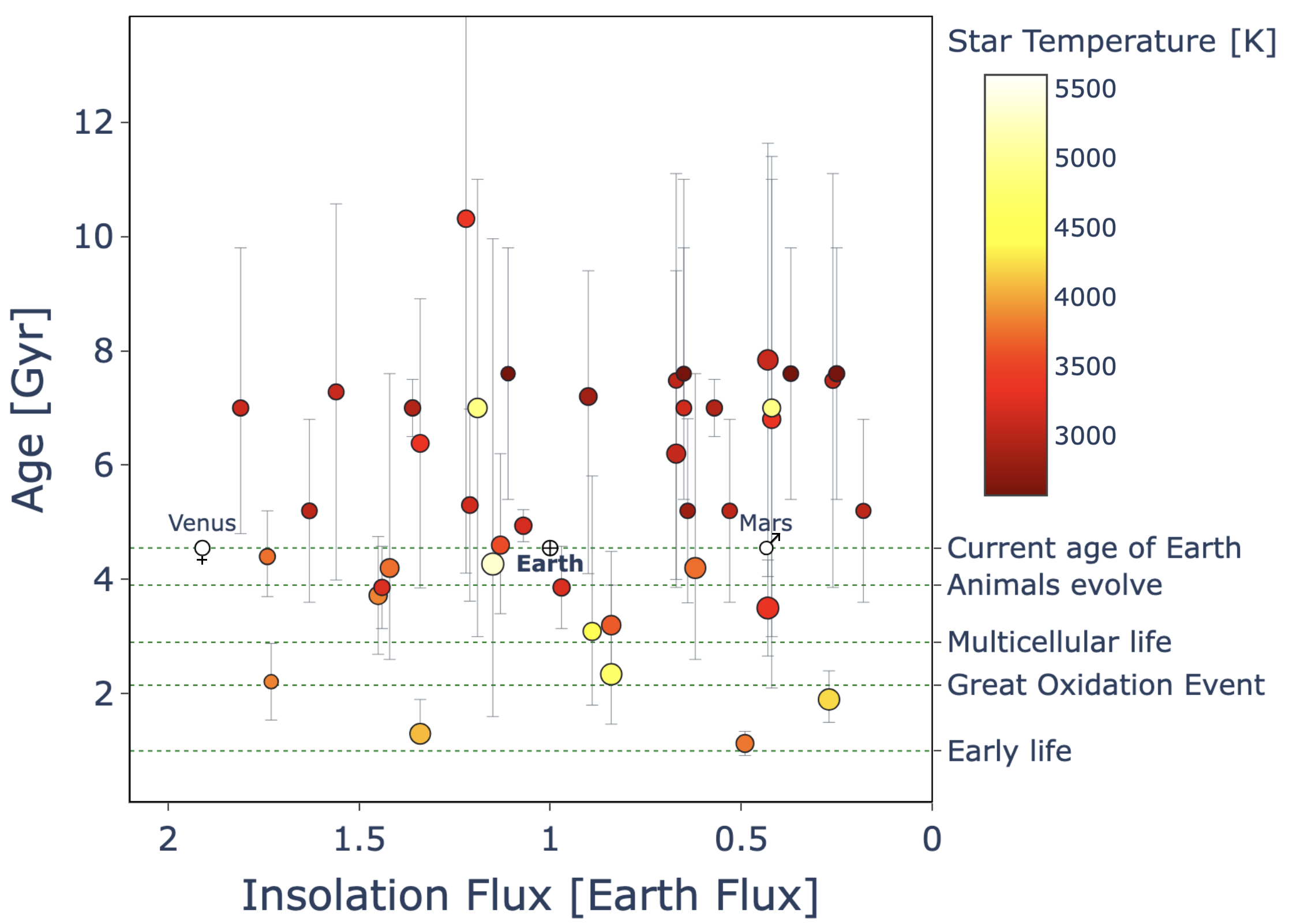}
    \caption{Tabulated age estimates for rocky HZ planets vs. stellar irradiation up to 2 times modern Earth's. Dot sizes correspond to planet radii or minimum masses and colours to host star $\bm{T}_\text{eff}$. Planets that require flux uncertainties to be in the HZ are included, but planets that could be rocky based on size uncertainty are excluded. Earth, Venus, and Mars are included for reference.}
    \label{fig:age}
\end{figure}

The host stars and rocky HZ exoplanets with the oldest estimated nominal ages (transiting and non-transiting) are shown in Table \ref{tab:age}. Among the 30 exoplanet host stars with age estimates, 17 stars (and their 24 exoplanets) have nominal ages older than Earth's, which would allow a similar or longer timescale for evolution on those exoplanets. High uncertainties in these age estimates are reflected in the large error bars in Table \ref{tab:age} and Figure \ref{fig:age}. 

Figure \ref{fig:age} plots tabulated age estimates versus $\bm{T}_\text{eff}$ of rocky HZ planets' host stars and provides the context of important milestones in the evolution of life on Earth, although the exact times of older milestones are debated. 
Fossils indicating that life had been established on Earth date back to about 1\,Gyr after Earth's formation (3.5\,Gyrs ago) \cite[e.g.,][]{Nutman2016}, but life might have already originated beforehand \cite[e.g.,][]{Bell2015}. 
The Great Oxidation Event about 2.45\,Gyr ago (about 2\,Gyr after Earth's formation) marked the accumulation of atmospheric O$_2$, as a result of oxygenic photosynthesis that had developed earlier \cite[e.g.,][]{Kasting2013}. 
Multicellular fossils date back to around 1.6\,Gyr ago (about 2.9\,Gyr after Earth's formation) \citep{Miao2024}. 
Land plants conquered the land masses at about 0.5\,Gyr ago (about 4\,Gyr after Earth's formation)\citep{Lenton2016}; 
dinosaurs roamed between about 250 to 66 million years ago; and humans evolved only recently on this large timescale, with the oldest Homo sapiens fossil claim about 300,000 years ago (\cite{Callaway2017}).

\section{Discussion} 

\subsection {Customizing the HZ planet list for specific criteria}
The reader can easily choose to create a subset of planets from our final lists, e.g., creating a list for the 3D HZ, or including planets that may be in the HZ due to the uncertainties in stellar measurement. The number of rocky exoplanets that might be in the empirical HZ increases by 9 if one includes the whole possible range of flux due to measurement uncertainties in stellar parameters of the host stars, which influence the incident flux. The list of rocky HZ planets then becomes 54 (32 of which transit). If we also include the measurement uncertainties on planet size, that number increases further to 73 (44 of which transit). 
%
Using the 3D inner limit rather than the empirical inner limit results in 24 rocky exoplanets in the 3D HZ. If one includes all rocky exoplanets that might be in the 3D HZ with measurement uncertainties in flux, that number increases to 41. 

In addition to the rocky HZ planet list, we create a full list of HZ exoplanets for observers interested in a wide range of planet types, using both i) host stars based on NEA with {\it Gaia} DR3 stellar parameter updates (NEA-{\it Gaia}) and ii) NEA-only stellar data (see discussion below on the difference between the lists due to differences in stellar data). Both lists are available online (\cite{bohl_zenodo}).

\subsection {Comparison to Earlier Work}
An influential 2023 paper \citep{hill2023} used solely NEA stellar data for their analysis of rocky planets in the HZ. To account for planets that had not been discovered at that time, we repeated the \cite{hill2023} analysis for NEA data only, and found the same results when assuming the same criteria for selection. Note that there were 7 planets, including one rocky planet, since identified as false positives in the \cite{hill2023} HZ planet list (HD~217850~b, HD~131664~b, GJ~832~c, HD~114613~b, KIC~5951458~b, BD-00~4475~b, HD~203473~b).


As this analysis shows, a small percentage of exoplanets can turn out to be false positives. To avoid false-positives, we chose to use the confirmed planet assignment on the NASA Exoplanet Archive in our analysis, but we note that four planets (Kepler-186~f, Kepler-452~b, and GJ~667~C~e \&~f) are flagged as controversial in the NEA. Those may turn out to be false positives. Their names are followed by an asterisk sign in our tables to indicate their controversial nature.

As shown in our comparison to the results from \citet{hill2023}, a few false positives can be found even in confirmed planet data. Before Kepler, transit surveys required independent confirmation through a series of follow-up observations to exclude false-positive scenarios such as stellar eclipsing binaries. However, only a small number of Kepler planets could be verified due to large observation time requirements either by RV or by transit timing variation in multiple-planet systems, prompting the development of probabilistic validation to demonstrate that all astrophysical false positive scenarios are negligibly likely \citep[for more details see e.g.,][and references therein]{Morton2016,Ross2024}. In addition, it can be difficult to separate RV signals caused by stellar activity from those resulting from orbiting planets, which highlights the importance of simultaneous photometric measurements to derive the stellar rotation periods and identify activity signals \citep[for more details see e.g.,][and references therein]{Vanderburg2016}.

%
To assess the potential habitability of eccentric exoplanets, \cite{hill2023} included planets that spend any time in the HZ. The percentages of each planet's orbit within the HZ, including uncertainties and without uncertainties, are both listed in the tables available at \cite{bohl_zenodo}, so the reader can also choose to use a smaller list of planets that spend 100\% of their time in the HZ.
We chose to follow \citet{Bolmont2016} in comparing the averaged flux for eccentric planets to the limits of the HZ. The averaged flux values account for the eccentricity of an HZ planet's orbit, assuming that an Earth-like atmosphere can buffer some orbital eccentricity, but only to a certain point. Note that due to using the averaged flux, our eccentricity inclusion is more conservative than including all planets that spend any time in the HZ, but less conservative than requiring 100\% time in the HZ.

\subsection {Exoplanet systems with large measurement uncertainties that might be in the HZ}
Stellar data is critical to assess which planets are in the HZ, because HZ limits depend on incident stellar irradiation. Whether or not planets could be rocky is also determined by stellar parameters, because planetary radius and mass are calculated in comparison to stellar parameters. In addition, all measurements have uncertainties; our analysis also identified the stars with measurement uncertainties that may allow for their planets to be located in the HZ. For these specific stars, observations to constrain these parameters and measurement uncertainties further can either include or exclude those exoplanets from the HZ. When we include the full range of possible uncertainties on the stellar parameters and eccentricity, the number of rocky planets that might be in the empirical HZ increases to 54 (5 of these 9 potentially additional rocky candidates in the HZ transit: Gliese 12 b, Kepler-399 d, Kepler-438 b, K2-72 c, and TOI-1452 b, while 4 do not transit: Kepler-138 e, Ross 128 b, Teegarden's Star b and d). Updated parameters for these systems will be able to characterize their incident irradiation further and exclude them from or include them in the empirical HZ. 

\subsection {Effect of the update of stellar data from Gaia DR3}
In this study, we have updated the NEA exoplanet host star parameters with {\it Gaia} DR3 for stars with RUWE <1.4, if data is available (hereafter referred to as NEA-{\it Gaia} data). The Gaia DR3 dataset provides a uniformly derived set of stellar parameters that is especially impactful for deriving broad population-level information. Thus, we have chosen to use the DR3 data when available. However, the NEA data is individually compiled for each star and can provide improved estimates for individual stars. Here we provide an analysis of the differences of the rocky HZ planet derived depending on whether one used the original NEA data or the updated GAIA DR3 stellar data.

Updating the host star properties influences not only the incident flux at the planet's position, but also where the HZ limits lie in terms of orbital distances for the system, because the parametrization of the HZ flux limits depends on $\bm{T}_{\text{eff}}$ \cite[e.g.,][]{Kasting1993}. Thus, some planets gain—and several others lose—their assignment as planets in the HZ due to the update to DR3 stellar data. Note that stellar radii changes due to DR3 updates also in turn change the associated planetary radii accordingly because a planet's size is measured as a fraction of the star's size; some planets fall below 2 Earth radii or exceed 2 Earth radii because of the update.

To identify the changes due to {\it Gaia} DR3 updates of stellar data, we ran an analysis on the full NEA-only data for comparison. While there is still a large overlap between rocky planets in the HZ, this comparison shows how critical stellar parameters are in determining which planets orbit in the HZ, as well as their planetary characteristics, since these are derived from stellar data.

\begin{table*}
    \centering 
    \begin{tabular}{lclccccccrcclcc} 
        \toprule
        Planet Name& Radius & Mass & {Flux} & {Min} & {Max} & {RV} & {3D} & {EM} & {TSM} & {$\theta$} & {Contrast} & { Age} & {d} & {$\bm{T}_\text{eff}$}\\
        & {R$_\oplus$} & {M$_\oplus$} & {$S_0$} & {$S_0$} & {$S_0$} & {S$_0$} & {$S_0$} & {$S_0$} & {} & {mas} & {} & {Gyr} & {pc} & {K}\\
        \midrule
        TRAPPIST-1 d & 0.79$^{\pm0.01}$ & 0.39$^{\pm0.01}$ & 1.11 & 1.03 & 1.21 & 1.47 & 0.77 & 0.20 & 22.75 & 1.78 & 1.71e-7 & 7.6$^{\pm2.2}$& 12.47 & 2566$^{\pm26}$\\
        TRAPPIST-1 e & 0.92$^{\pm0.01}$ & 0.69$^{\pm0.02}$ & 0.65 & 0.51 & 0.82 & 1.47 & 0.77 & 0.20 & 17.71 & 2.34 & 1.35e-7 & 7.6$^{\pm2.2}$ & 12.47 & 2566$^{\pm26}$\\
        TRAPPIST-1 f & 1.01$^{+0.01}$ & 1.04$^{\pm0.03}$ & 0.37 & 0.34 & 0.4 & 1.47 & 0.77 & 0.20 & 15.07 & 3.09 & 1.00e-7 & 7.6$^{\pm2.2}$ & 12.47 & 2566$^{\pm26}$\\
        TRAPPIST-1 g & 1.13$^{+0.02}_{-0.01}$ & 1.32$^{\pm0.04}$ & 0.25 & 0.23 & 0.27 & 1.47 & 0.77 & 0.20 & 13.55 & 3.76 & 7.92e-8 & 7.6$^{\pm2.2}$ & 12.47 & 2566$^{\pm26}$\\
        LHS 1140 b & 1.73$^{\pm0.02}$ & 5.60$^{\pm0.19}$ & 0.43 & 0.38 & 0.49 & 1.50 & 0.81 & 0.21 & 8.93 & 6.31 & 4.56e-8 & 7.84$^{\pm3.8}$ & 14.99 & 3096$^{\pm48}$\\
        $\cdots$ & $\cdots$ & $\cdots$ & $\cdots$ & $\cdots$ & $\cdots$ & $\cdots$ & $\cdots$ & $\cdots$ & $\cdots$ & $\cdots$ & $\cdots$ & $\cdots$ & $\cdots$ & $\cdots$\\
        Proxima Cen b & & 1.06$^{\pm0.06}$ & 0.54 & 0.33 & 0.82 & 1.49 & 0.79 &  0.21 & 169.21 & 37.26 & 5.99e-8 & 5.2$^{\pm1.61}$ & 1.30 & 2900$^{\pm100}$\\
        Wolf 1061 c & & 3.41$^{+0.43}_{-0.41}$ & 1.34 & 0.98 & 1.83 & 1.51 & 0.83 & 0.22 & 27.31 & 20.67 & 3.35e-8 & 6.38$^{\pm2.5}$ & 4.31 & 3342$^{\pm49}$ \\
        GJ 273 b &  & 2.89$^{+0.27}_{-0.26}$ & 1.22 & 0.94 & 1.56 & 1.52 & 0.84 & 0.22 & 33.69 & 15.38 & 2.93e-8 & 10.31$^{\pm6.2}$ & 5.92 & 3382$^{\pm49}$ \\ 
        Ross 128 b &  & 1.40$^{\pm0.21}$ & 1.47 & 1.17 & 1.87 & 1.50 & 0.82 & 0.21 & 60.80 & 14.80 & 6.6e-8 & 7.3$^{\pm3.3}$ & 3.37 & 3192$^{\pm60}$ \\  
        GJ 1002 c &  & 1.36$^{\pm0.17}$ & 0.26 & 0.21 & 0.32 & 1.49 & 0.80 & 0.21 & 36.31 & 15.22 & 2.97e-8 & 7.48$^{\pm3.6}$ & 4.85 & 3024$^{\pm52}$ \\  
        GJ 1061 d & & 1.64$^{+0.24}_{-0.23}$ & 0.57 & 0.45 & 0.72 & 1.49 & 0.80 & 0.21 & 47.60 & 14.70 & 6.13e-8 & 7.0$^{\pm0.5}$ & 3.67 & 2953 $^{\pm98}$\\
        \bottomrule
    \end{tabular}
    \caption{Sample table of the Catalog of Habitable Zone Rocky Planets based on NEA data only (without {\it Gaia} DR3 updates). Here planets are sorted as in Table~\ref{tab:tableRockyHZ} by whether their nominal values place them as rocky HZ planets, then by descending TSM values for those that transit (above dotted row), and by angular separation ($\theta$) for contrast ratios above $10^{-8}$ for direct imaging of those that do not transit (below dotted row). 
    Incident flux, with the minimum and maximum value (Flux, Min, Max), two inner limits of the HZ—the empirical Recent Venus (RV) and a 3D model (3D) limit—and the outer empirical Early Mars (EM) limit are provided in units of modern Earth flux ($S_0$). The maximum and minimum possible stellar flux reaching each planet is calculated based on measurement uncertainties in stellar temperature ($\bm{T}_\text{eff}$), measurement uncertainties in semi-major axis, and nominal eccentricity. 
    The full table is available at \citet{bohl_zenodo}.}
    \label{tab:tableRockyHZNEA}
\end{table*}

NEA-only stellar data without including measurement uncertainties finds 42 rocky planets in the empirical HZ  (see Table \ref{tab:tableRockyHZNEA}), compared to 45 with DR3 updates (see Table \ref{tab:tableRockyHZ}).

\textbf{\begin{table}
    \centering 
    \begin{tabular}{ll} 
        \toprule
        Category & Planet Names \\
        \midrule
         \multirow{10}{*}{Overlap (38)}
        & GJ 251 c, GJ 273 b, GJ 1002 b \& c, GJ 1061 c \& d, \\
        & GJ 3323 b, K2-3 d, K2-72 e, K2-288 B b, \\
        & Kepler-62 e \& f, Kepler-186 f, Kepler-296 e \& f, \\
        & Kepler-442 b, Kepler-452 b, Kepler-1229 b, \\
        & Kepler-1410 b, Kepler-1544 b, Kepler-1649 c, \\
        & Kepler-1652 b, L 98-59 f, LHS 1140 b, LP 890-9 c, \\
        & Proxima Cen b, Ross 508 b, TOI-700 d \& e, \\
        & TOI-715 b, TOI-1266 d, TRAPPIST-1 d \& e \& f \& g, \\
        & Teegarden's Star c, Wolf 1061 c, Wolf 1069 b \\
        \midrule
        NEA-Gaia (+7)& GJ 667 C c \& e* \& f*, GJ 682 b, \\
        not & K2-239 d , Kepler-1606 b, \\ 
        NEA-only  & Kepler-441 b,  \\ 
        \midrule
        NEA-Only (+4)& Kepler-1638 b, Kepler-283 c, Kepler-440 b,\\
        not NEA-Gaia & Ross 128 b\\ 
        \bottomrule
    \end{tabular}
    \begin{tablenotes}
	\small {
	\item{* Indicates controversial planets in the NEA (see discussion).}}
    \end{tablenotes}
    \caption{Comparison of rocky planets in the HZ using NASA Exoplanet Archive data only (NEA-only) vs. updated stellar data based on {\it Gaia} DR3 parameters for RUWE <1.4 (NEA-{\it Gaia}). 
    }
    \label{tab:tableNEAvsGaia}
\end{table}}
38 rocky HZ planets overlap in both lists (NEA-only: 42, NEA-{\it Gaia}: 45)(see Table \ref{tab:tableNEAvsGaia}).
7 planets enter the HZ when updating stellar data with DR3: 4 of these 7 planets are missing stellar radii in the NEA (GJ~667~C~c \&~e* \&~f*, GJ~682~b), 2 others change stellar $\bm{T}_\text{eff}$ and stellar radius (K2-239~d, Kepler-441~b), and 1 planet radius changes (Kepler-1606~b) due to a stellar radius update, moving it below 2 Earth radii.
4 planets in the NEA-only list are not in the updated rocky planet NEA-{\it Gaia} HZ list (Kepler-1638~b, Kepler-283~c, Kepler-440~b, Ross~128~b). 3 of these host stars have different $\bm{T}_\text{eff}$ and stellar radius (Kepler-1638~b, Kepler-440~b, Ross~128~b), and one planet's radius changed from rocky to not with a stellar radius update (Kepler-283 c).

While there is a large overlap between rocky planets in the HZ, this comparison shows how critical the stellar parameters are in determining which planets orbit in the HZ and their characteristics, such as radius and minimum mass, as those are derived based on stellar data. Especially for planets in one but not both lists, a more in-depth study of stellar parameters is important to assess whether these worlds could be potentially habitable. Exploration of {\it Gaia} stellar data for exoplanet host stars is underway to constrain their stellar parameters further \citep[see e.g.,][]{laverny2025}.
%

%

\subsection {Stellar Activity}

Multiple M dwarfs in our list have recorded instances of stellar activity. 5 stars and their 7 planets (out of 45 rocky HZ planets) have at least one recorded flare in TESS data \citep{Yang2023, Pietras2022}: GJ~1061, GJ~273, GJ~3323, L~98-59, and Proxima~Cen. Some teams \cite[e.g.,][]{Atkinson2024} have calculated a planet's habitability based on the activity of their host star, which allocates an Alfvén Surface Habitability Criterion below 1 for the rocky HZ planets GJ~1002~b and c, GJ~273~b, GJ~3323~b, Teegarden's~Star~c, TRAPPIST\nobreakdash-1~d, Proxima~Cen~b, and Wolf~1061~c, and Wolf~1069~b. These planets may arguably require a strong magnetic field to protect surface life or otherwise may experience accelerated atmospheric loss. In addition, UV flux limits may arguably influence surface habitability in parts of the HZ \citep{Spinelli2024}, especially for M dwarfs $< 2800$\,K. 
However, note that these concerns mostly address surface habitability, and do not exclude habitability on such worlds for either subsurface or radiation-adapted life \cite[e.g.,][]{O'Malley-James2017}. In addition, UV levels on the surface are not only determined by O$_3$, but by additional atmospheric molecules such as CO$_2$ and H$_2$O, which block UV radiation from reaching the ground shortwards of 200\,nm. This could provide effective UV protection above 200\,nm for exoplanets, e.g., with Archean, CO$_2$-, or H$_2$O-rich atmospheres \citep{O'Malley-James2017}, making such planets interesting targets to assess the influence of UV radiation on their surface habitability.

\subsection {Target selection for transit observations, direct imaging, and light curve measurements for missions in the design stage, such as HWO and LIFE}
We analysed all the NEA exoplanets known to date, identifying the rocky HZ planets so that observers and modelers can choose rocky HZ exoplanets based on their priorities, e.g., J magnitude, estimated age of the system, apparent angular separation, contrast ratio, and TSM, to name a few of the possibilities. 
For transit observations, we chose to use the TSM as described earlier in Methods, which allows observers to prioritize planets for transmission spectroscopy. Our rocky HZ target list can help observers prioritize targets for current and future JWST observations. In addition, in about 2030, the ELT will become available for science observation; observers can use this target list to plan observations and prioritize exoplanets for spectroscopic follow-up.

Specific instruments for direct observations have different limitations in terms of contrast ratio and apparent angular separation. Thus, we chose a generic contrast ratio limit above $10^{-8}$ for the targets shown in Table \ref{tab:contrast-tsm-ang}, so that the target list for direct imaging presented in this paper does not use specific instrument requirements. Table \ref{tab:tableRockyHZ} provides a wide range of curated parameters for the rocky HZ planets and their host stars, allowing observers to create their own prioritized target list by identifying which of these HZ rocky planets are the best targets for direct imaging, given their specific instrument specifications.

As another example of how this target list could be used, we select the planned NASA HWO mission concept that is currently undergoing trade-off studies of three baseline design ideas. While HWO is currently not designed to observe any of the known HZ rocky planets, and thus there is currently no overlap between its target stars \cite{mamajek2023} and the known rocky HZ planets we identified here, this might inspire creative ideas for updating its mission designs if known exoplanets are considered a priority. Another example is the European LIFE mission concept \cite[e.g.,][]{quanz2022} that is currently in an early design phase and has not yet released a final list of target stars. Thus, this list of rocky planets in the HZ can provide interesting known exoplanet targets for mission design and any trade-off studies.

\section{Conclusion} 
To assess the limits of surface habitability, it is critical to characterize rocky exoplanets in the HZ: observations of known rocky exoplanets on the edges of the HZ can now empirically explore these boundaries. 

In this paper, we analysed the data from Gaia DR3 and the NASA Exoplanet Archive (NEA) of all known exoplanets, identifying those that can i) can probe of the limits of habitability orbiting at the edges of the HZ, ii) provide similar irradiation environments to modern Earth, iii) explore the effect of eccentricity on habitability, iv) are potentially evolved rocky worlds, and v) present targets to prioritize for transmission observations, light curve measurements, and direct imaging.
We presented a target list of 45 rocky exoplanets in the empirical HZ (27 transiting) and 24 in a narrower 3D HZ (15 transiting). We also discussed the differences in targets when using NEA-only data compared to updating the host star data with GAIA DR3. 
We also provided the theoretical limits for the empirical HZ and a 3D-HZ for each system, and identified the oldest HZ rocky worlds based on data from the NEA and complementary literature data.

We compared the demographics of the rocky HZ planets with the full catalog of exoplanets in the NASA Exoplanet Archive (NEA), and calculated the transmission spectroscopy metric as a guide for transit observations, as well as the contrast ratio and apparent angular separation as a guide for secondary eclipse and direct imaging observations. 

We also tabulate age estimates for the rocky planets in the HZ to compare their possible stage of evolution to life on Earth. \textbf{23} exoplanets have nominal ages older than Earth's, which could allow a similar or longer timescale for evolution on those exoplanets. 

We identified the oldest rocky HZ exoplanets (Table \ref{tab:age})
and rocky HZ planets with the highest TSM, angular separation, and contrast ratio as priority targets for transit observations and for direct imaging and light curve measurements (Table \ref{tab:contrast-tsm-ang}).

We provided the best transiting and non-transiting planets to test the limits of surface habitability near the inner and outer regions of the empirical HZ (Table \ref{tab:tableLimits}).
%
In addition to probing the edges of habitability, we showed in the same table which rocky planets receive incident stellar flux comparable to Earth's, making them very interesting targets for further observations.
%
We also highlighted eccentric rocky HZ planets that can assess the influence of eccentricity on habitability (Table \ref{tab:tableecc}).

The resulting planetary target characteristics allow observers to shape and optimize their search strategies with space- and ground-based telescopes---such as the James Webb Space Telescope, extremely large telescopes (ELTs), and concepts like the Habitable Worlds Observatory (HWO) and LIFE---and design new observing strategies and instruments to explore these intriguing worlds, addressing the question of the limits of surface habitability on exoplanets.

\section*{Acknowledgments}
This research has made use of the NASA Exoplanet Archive, which is operated by the California Institute of Technology, under contract with the National Aeronautics and Space Administration under the Exoplanet Exploration Program.

This work has made use of data from the European Space Agency (ESA) mission
{\it Gaia} (\url{https://www.cosmos.esa.int/gaia}), processed by the {\it Gaia}
Data Processing and Analysis Consortium (DPAC,
\url{https://www.cosmos.esa.int/web/gaia/dpac/consortium}). Funding for the DPAC
has been provided by national institutions, in particular the institutions
participating in the {\it Gaia} Multilateral Agreement.

We thank Eric Mamajek, Jessie Christiansen, Rebecca Payne, and the anonymous referee for helpful comments and discussions.

\section*{Data Availability}
The final data (with values from the NASA Exoplanet Archive, from {\it Gaia} DR3, and those calculated in this work) are available on zenodo (\url{https://doi.org/10.5281/zenodo.18134528}) (\cite{bohl_zenodo}).

\bibliography{citations}

\bibliographystyle{mnras}

\end{document}